\documentstyle[aps,prl,multicol,psfig]{revtex}

\begin{document}
\draft

\title{The role of electronic correlation in the Si(100) reconstruction:\\
a quantum Monte Carlo study}
\author{Sorcha B. Healy, Claudia Filippi}
\address{Physics Department, National University of Ireland, Cork, Ireland}
\author{P. Kratzer, E. Penev, and M. Scheffler,}
\address{Fritz-Haber-Institut der Max-Planck-Gesellschaft, Faradayweg 4-6,
D-14195 Berlin-Dahlem, Germany}
\date{\today}
\maketitle

\begin{abstract}
Recent low-temperature scanning tunneling experiments have challenged the
generally accepted picture of buckled silicon dimers as the ground state
reconstruction of the Si(100) surface.
Together with the symmetric dimer model of the surface suggested by quantum
chemistry calculations on small clusters, these findings question our
general understanding of electronic correlations at surfaces and its proper
description within density functional theory.
We present quantum Monte Carlo calculations on large cluster models of the
symmetric and buckled surface, and conclude that buckling remains energetically 
more favorable even when the present-day best treatment of electronic
correlation is employed.
\end{abstract}


\begin{multicols}{2}

Despite extensive experimental and theoretical investigation, the
nature of the reconstruction of the Si(100) surface is still subject
to debate.
While this surface is of technological relevance because of its use in 
the fabrication of electronic devices, determining its ground state structure 
is important as a test of our general understanding of the role of electronic 
correlation at surfaces.

Scanning tunneling microscopy (STM) experiments indicate that Si(100)
reconstructs in rows of silicon dimers~\cite{Hamers}.
At room temperature, most dimers appear symmetric due to their dynamical
flipping motion, and only dimers close to defects are pinned in a buckled
configuration~\cite{Wolkow}. The number of symmetric dimers decreases below
120~K and dimers buckle alternately within each row, with the formation
of $p(2 \times 2)$ or $c(4 \times 2)$ domains corresponding to adjacent 
rows in identical or opposite orientations~\cite{Wolkow,Tochihara}.  
Experimentally, the $c(4\times 2)$ reconstruction (Fig.~\ref{fig1}) 
was accepted as the lowest energy structure.

\noindent
\begin{minipage}{3.375in}
\begin{figure}
\noindent
\centerline{\psfig{figure=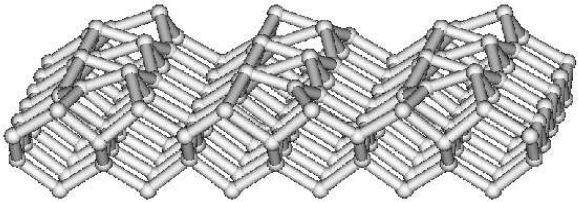,width=8.0cm}}
\vspace*{0.3cm}
\caption{Model of the silicon (100) surface in the $c(4\times 2)$ reconstruction.}
\label{fig1}
\end{figure}
\vspace{1ex}
\end{minipage}
This picture is now being challenged by new experimental work.
A series of low temperature STM studies have recently reported that,
while the $c(4\times 2)$ structure is observed below 120~K, further
cooling below 20~K causes the dimers to appear again symmetric.
Yokoyama and Takayanagi~\cite{Yokoyama} argue that this is a dynamical
phenomenon caused by a lowering of the potential energy barrier between
the two buckled configurations, which allows the dimers to resume the
flip-flop motion characteristic of room temperature.  In contrast,
Kondo {\it et al.}~\cite{Kondo} claim that the observed symmetric
dimers are static since their images do not exhibit the noise
associated with the flipping motion observed in the same sample
at 110 K.
However, evidence has also been produced that, at low temperatures, 
the rate of dimer flipping is significantly affected by the amount of 
tunneling current~\cite{Mitsui}, casting doubts on the conclusions 
drawn by Kondo.

To date, theoretical work also remains divided on the issue.  
Density functional theory (DFT) calculations on slab 
geometries~\cite{Dabrowski1} and large cluster 
models~\cite{Penev,Konecny,Yang} favor a buckled reconstruction.
$GW$ calculations~\cite{Northrup,Rohlfing} on buckled geometries show good
agreement with the measured dispersion of surface band states~\cite{Johansson}, 
and also a surface core-level shift analysis~\cite{pehkle93}
supports the DFT finding. On the other hand, multiconfiguration 
self-consistent field (MCSCF) and configuration interaction
(CI)~\cite{Radeke1,Paulus,Jing,Shoemaker}
calculations on small clusters find the symmetric reconstruction to
have the lowest energy.
These quantum chemistry techniques emphasize different components of 
electronic correlation than DFT: static correlation, arising from  
near-degeneracy of molecular orbitals, is most effectively described by 
a linear combination of low-lying determinants as in a MCSCF 
calculation, while dynamical correlation, given by short-range electronic 
screening, is adequately treated in DFT.

The buckling of the Si dimers on the Si(100) surface is one of the hard problems
of many-body physics at surfaces, because it involves subtle aspects of
electronic correlation~\cite{Artacho}.
Since the symmetric surface dimer has two dangling bonds, each filled with
one electron, it is in a bi-radical state, and {\em static} correlation must
be properly taken into account. The buckled state, on the other hand, is
thought to be stabilized by a rehybridization of the dangling orbitals, 
accompanied by charge transfer: the lower Si atom approaches $sp^2$-like 
hybridization and its charge is depleted in favor of the higher Si atom 
in the dimer.
Thus, the net stability of the buckled configuration depends on the degree
to which the repulsive Coulomb interaction of the two electrons in the
upper Si electronic state is dynamically screened.
This is where {\em dynamical} correlation enters into the problem.

In this Letter, we use quantum Monte Carlo (QMC) methods to find accurate
energy differences between the symmetric and buckled reconstructions for
large cluster models of the surface. Unlike the other theoretical
methods previously used for this problem, QMC has the advantage that it can
be applied to sufficiently large systems and still provide an accurate
description of both dynamical {\it and} static electronic
correlation~\cite{qmc}.
We find that dimer-dimer interactions are important and sufficiently large
clusters must be used to adequately model the Si(100) surface.
Our many-body calculations conclusively show that the ground state of the
Si(100) surface is a buckled reconstruction, and that the trend with respect
to cluster size found in DFT calculations is correct.

{\it Cluster models of Si(100).}
While methods based on DFT are known to usually give a good description of
structural and elastic properties (e.g. surface lattice constant and elastic
interaction between the Si dimers), it is unclear whether they can adequately 
represent the subtle aspects of electronic correlation at the Si(100) surface.
It is therefore appropriate to use QMC on geometries obtained from DFT 
calculations to assess whether a more accurate treatment of
electronic correlation can fundamentally change the picture.
Here, we choose to address this issue by performing calculations for
clusters that mimic the surface geometry.

In identifying appropriate cluster shapes for Si(100), one must
recognize that interactions are negligible between neighboring
dimer rows, while they are substantial between dimers in the same row.
This is apparent from the small and the large dispersion of the surface
band states along the respective directions, $\Gamma-J$ and
$\Gamma-J'$~\cite{Rohlfing}.
Moreover, the $p(2 \times 2)$ and $c(4 \times 2)$ reconstructions are 
found energetically quite close in experiments and calculations~\cite{Inoue} 
(within 2 meV), demonstrating that dimer rows are weakly interacting.

Therefore, the surface can be modeled with clusters containing only a 
single row of dimers.  Such clusters with one, two and three dimers are
Si$_9$H$_{12}$, Si$_{15}$H$_{16}$ and Si$_{21}$H$_{20}$, previously
also used in Refs.~\cite{Penev,Konecny,Yang,Radeke1,Paulus,Jing,Shoemaker}.
They represent a four layer cut of the Si(100) surface with all but the 
surface atoms terminated with hydrogens to passivate dangling bonds.
In Fig.~\ref{fig2}, we show the Si$_{15}$H$_{16}$
and Si${}_{21}$H${}_{20}$ clusters both in the buckled and symmetric
reconstructions.

Most CI calculations have been performed on the smallest cluster
Si$_9$H$_{12}$ which is found to be symmetric~\cite{Radeke1,Jing}.
Also for the Si$_{15}$H$_{16}$ cluster, Paulus~\cite{Paulus} 
claims a symmetric ground state but the level of the CI calculation
is not specified and the energetics are not explicitly given.
 
The Si$_9$H$_{12}$ cluster is however too small to draw conclusions on the
real surface, as became clear from recent DFT calculations~\cite{Penev}.
For Si$_9$H$_{12}$, approximate density functionals yield conclusions
similar to CI, in that buckling is either energetically unfavorable or
marginally preferred by less than 4 meV.
However, as the number of dimers in the cluster increases, buckling
becomes favorable and the optimal buckling angle increases for all the
functionals. For the three-dimer cluster Si$_{21}$H$_{20}$, the energy gain
per dimer is between 0.15-0.20 eV, depending on the functional used, and
in agreement with the slab results to better than 0.05 eV.
This suggests that it is possible to infer the behavior of the Si(100) 
surface from the three-dimer cluster.

\noindent
\begin{minipage}{3.375in}
\begin{figure}[hbt]
\centerline{\psfig{figure=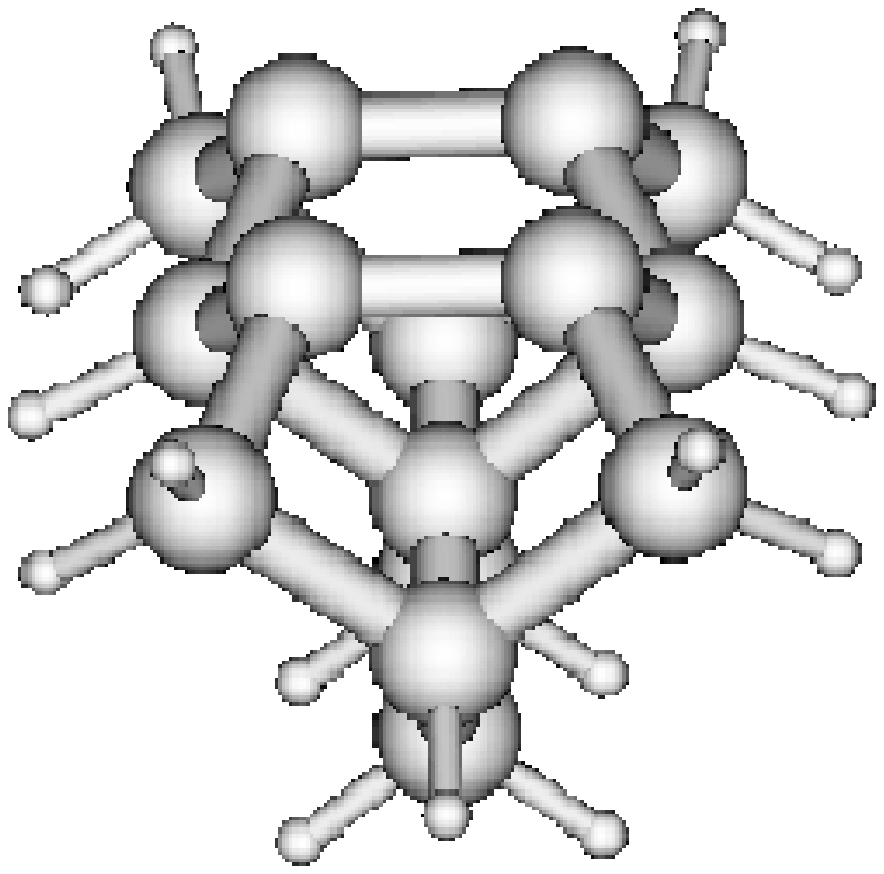,width=3.0cm}
\psfig{figure=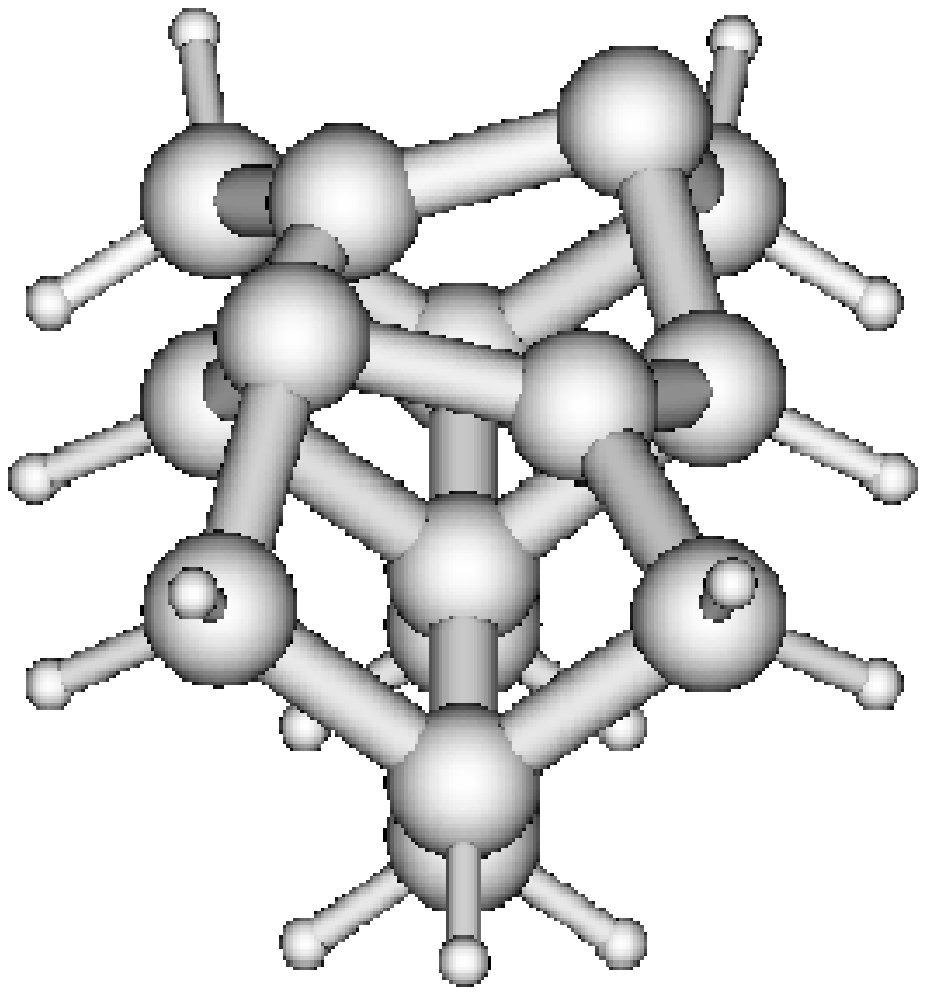,width=3.0cm}}
\vspace{10pt}
\centerline{\psfig{figure=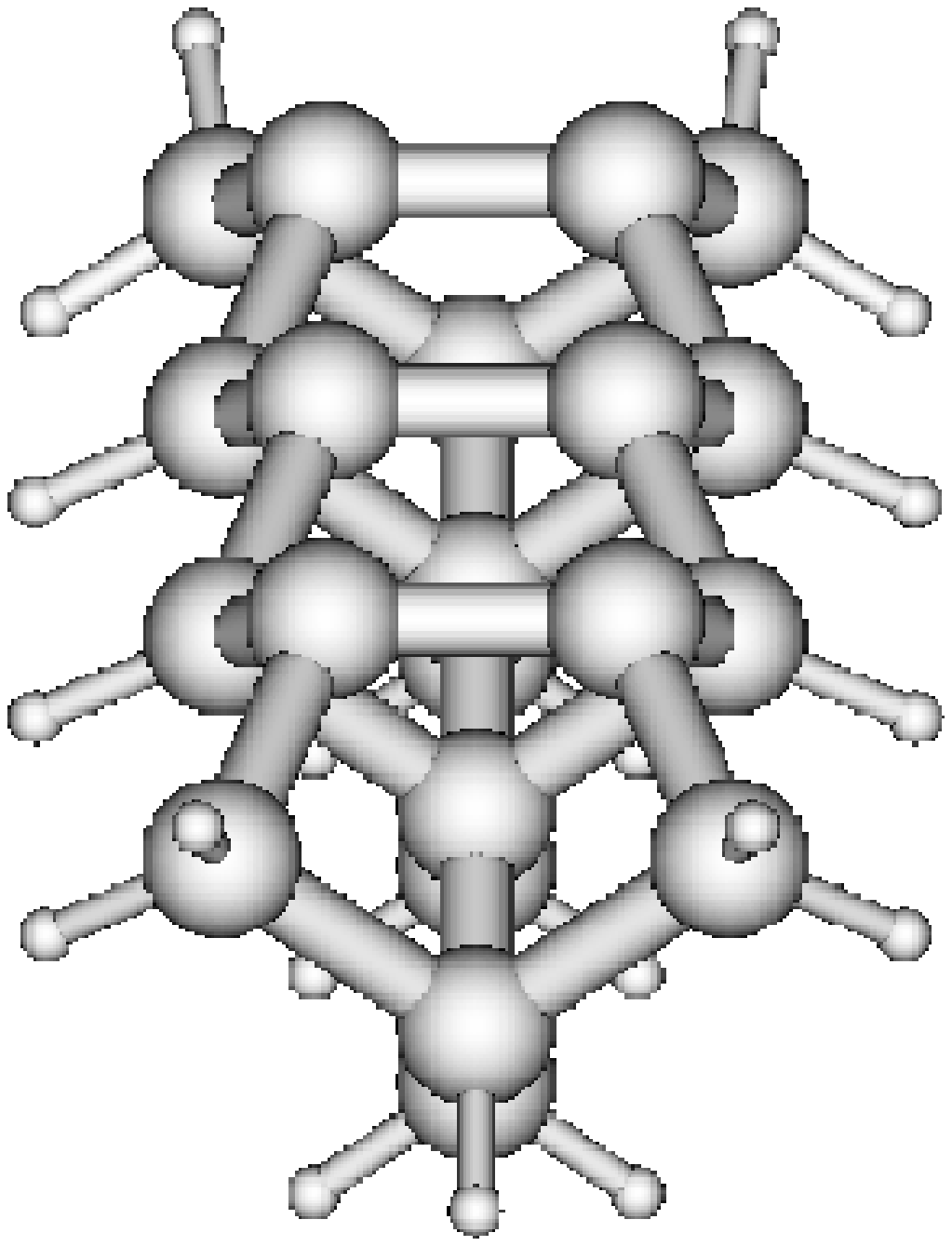,width=3.0cm}
\psfig{figure=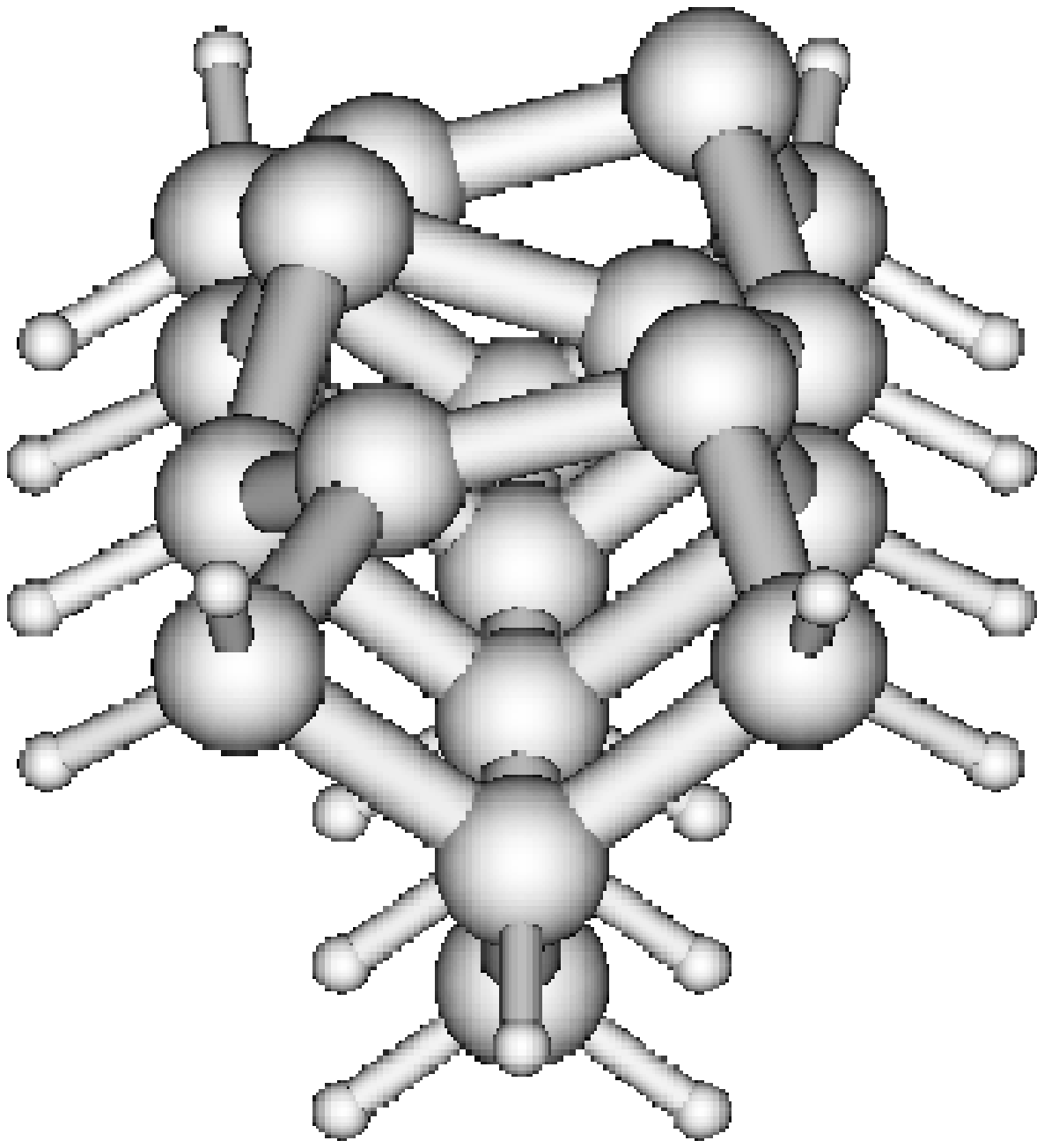,width=3.0cm}}
\vspace*{0.3cm}
\caption{Si$_{15}$H$_{16}$ (upper figure) and Si$_{21}$H$_{20}$ (lower
figure) cluster models of the Si(100) surface. Both the symmetric (left) 
and buckled (right) reconstructions are shown. The bigger atom is silicon 
and the smaller one is hydrogen.}
\label{fig2}
\end{figure}
\vspace{1ex}
\end{minipage}

Here, we perform QMC calculations on the clusters with two and three
dimers, Si$_{15}$H$_{16}$ and Si$_{21}$H$_{20}$.
For both clusters, we use the geometries optimized within DFT using
the PW91 functional~\cite{PW91}.
For details on the construction and the geometry of the clusters,
see Ref.~\cite{Penev}.

{\it Silicon pseudopotential.}
For the silicon atom, we use a norm-conserving {\it sp}-non-local
pseudopotential for the ten core electrons. The pseudo-potential was
generated in an all-electron Hartree-Fock calculation for the Si atom.
It was tested within QMC to calculate binding energy and bond length of 
Si$_2$ which were found in excellent agreement with experiment. The 
transferability of the pseudopotential with respect to all electron 
calculations was checked with Hartree-Fock (HF) and B3LYP~\cite{B3LYP} 
for larger silicon clusters.

{\it Quantum Monte Carlo methods.}
The form of the many-body wave function used in QMC calculations
efficiently describes the static part of electronic correlation by the
use of a linear combination of Slater determinants, as well as its
dynamical component by introducing a positive Jastrow correlation factor
(modified from Ref.~\cite{Filippi} to deal with pseudo-atoms):
\begin{equation}\label{wave function}
\Psi=\sum_{n} d_n D_n^{\uparrow} D_n^{\downarrow}
\prod_{\alpha i j}J\left(r_{ij},r_{i\alpha},r_{j\alpha}\right)\,.
\end{equation}
${\rm D}^\uparrow_n$ and ${\rm D}^\downarrow_n$ are Slater determinants
of single particle orbitals for the up and down electrons, respectively,
and the orbitals are represented using atomic Gaussian basis~\cite{basis}.
The Jastrow factor correlates pairs of electrons {\it i} and {\it j} with
each other, and with every nucleus $\alpha$, and different Jastrow
factors are used to describe the correlation with a hydrogen and
a silicon atom.

The determinantal part of the wave function is generated within HF or
MCSCF, using the quantum chemistry package GAMESS~\cite{GAMESS}.
As active orbitals in the MCSCF, we choose the occupied bonding and
the unoccupied antibonding $\pi$ orbitals, which are shown for the
Si$_{15}$H$_{16}$ symmetric cluster in Fig.~\ref{fig3}.

\noindent
\begin{minipage}{3.375in}
\begin{figure}[hbt]
\noindent
\centerline{\psfig{figure=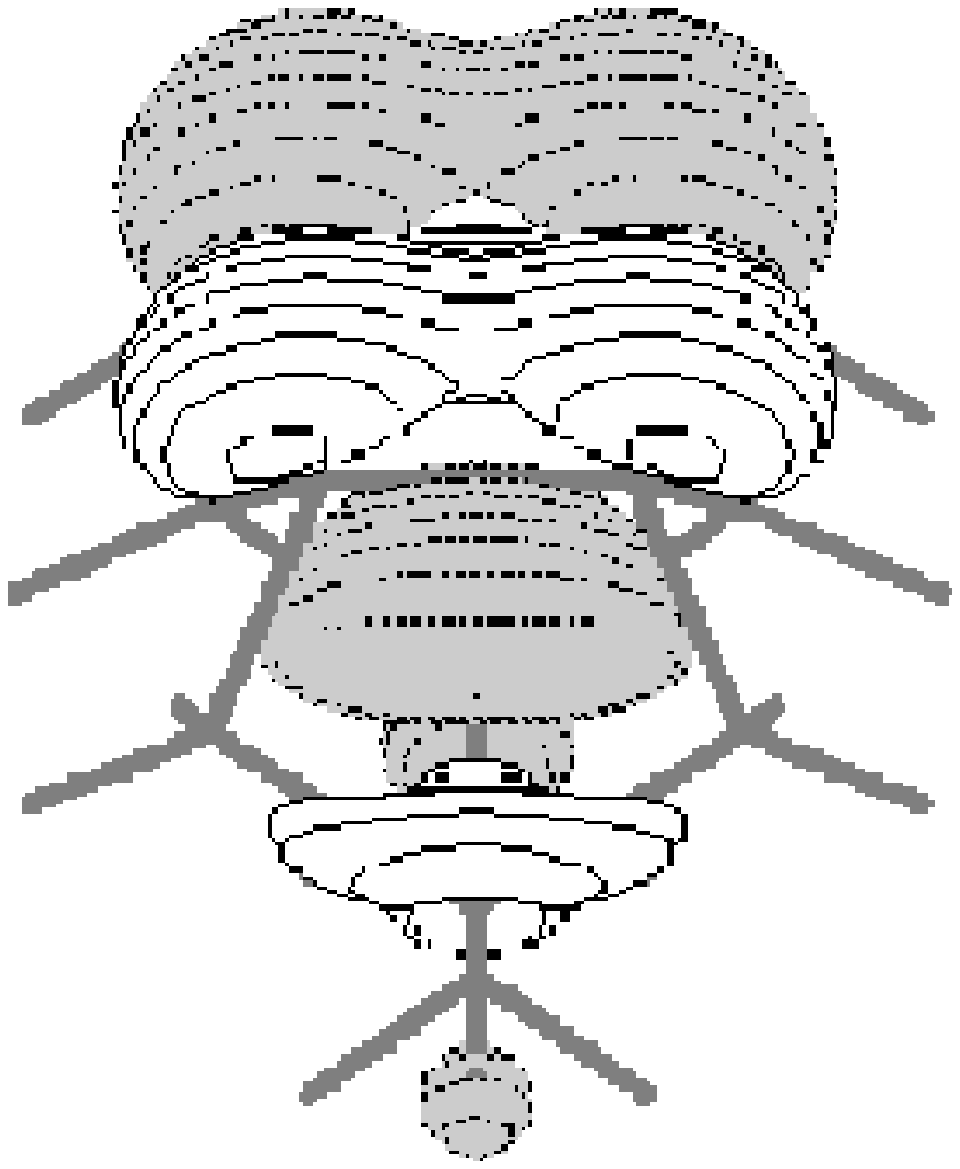,width=3.0cm}
\psfig{figure=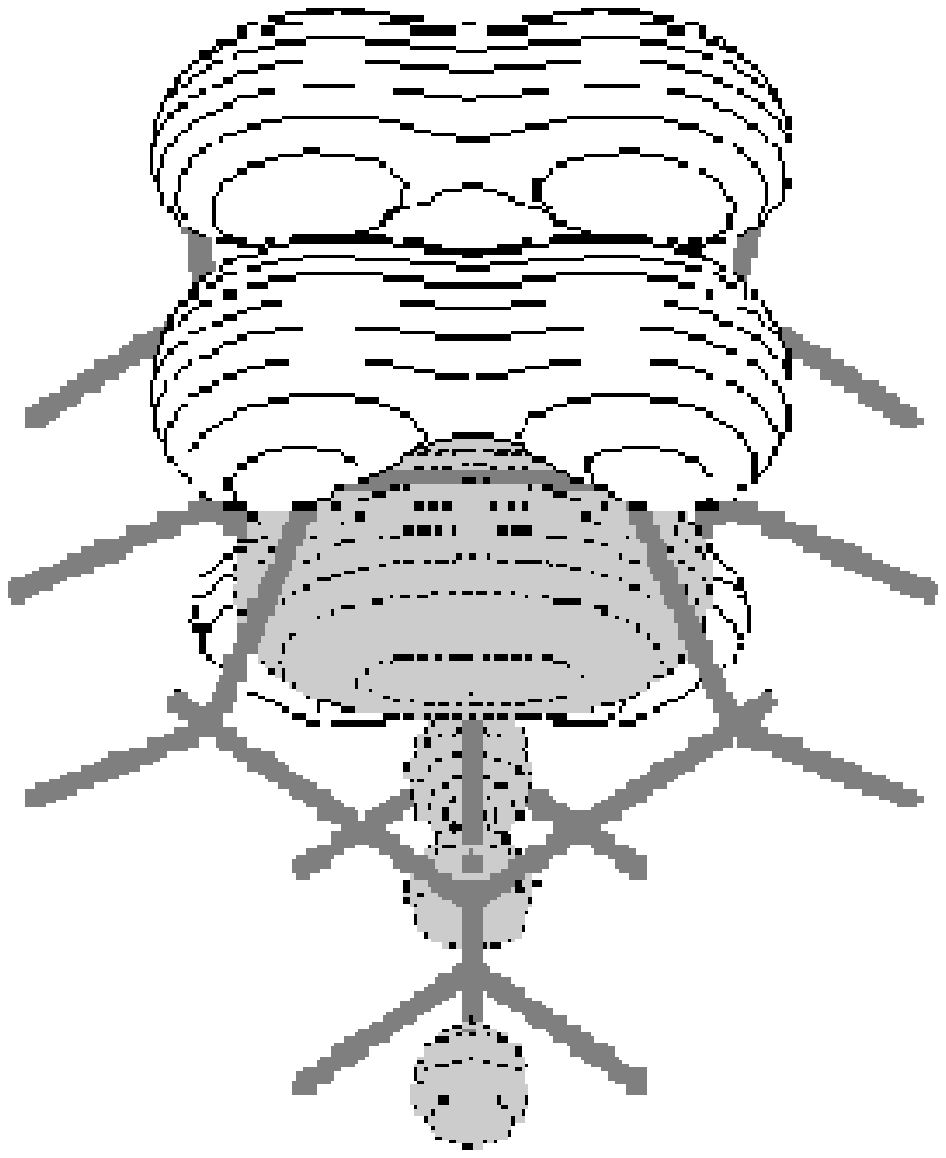,width=3.0cm}}
\vspace{10pt}
\centerline{\psfig{figure=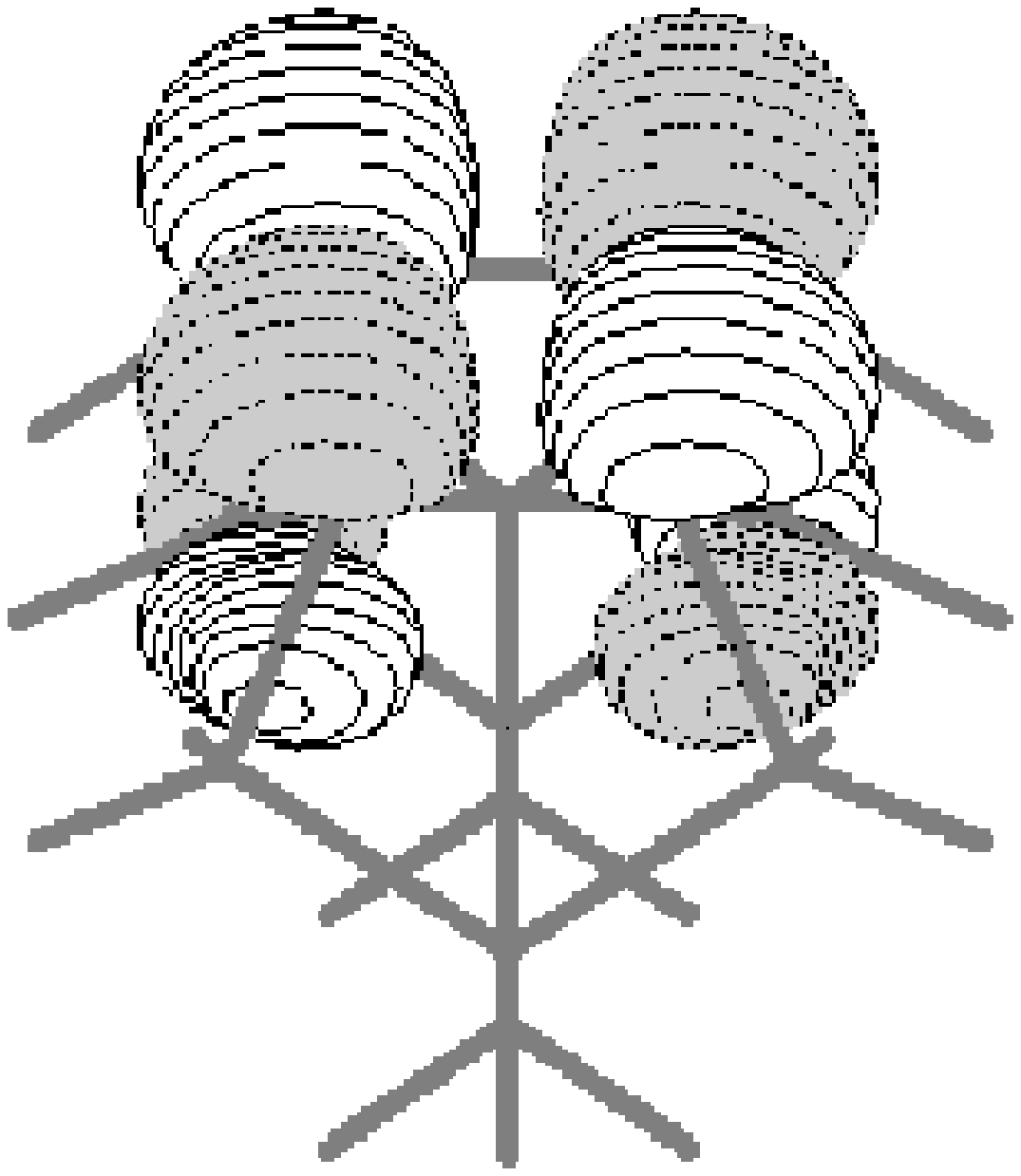,width=3.0cm}
\psfig{figure=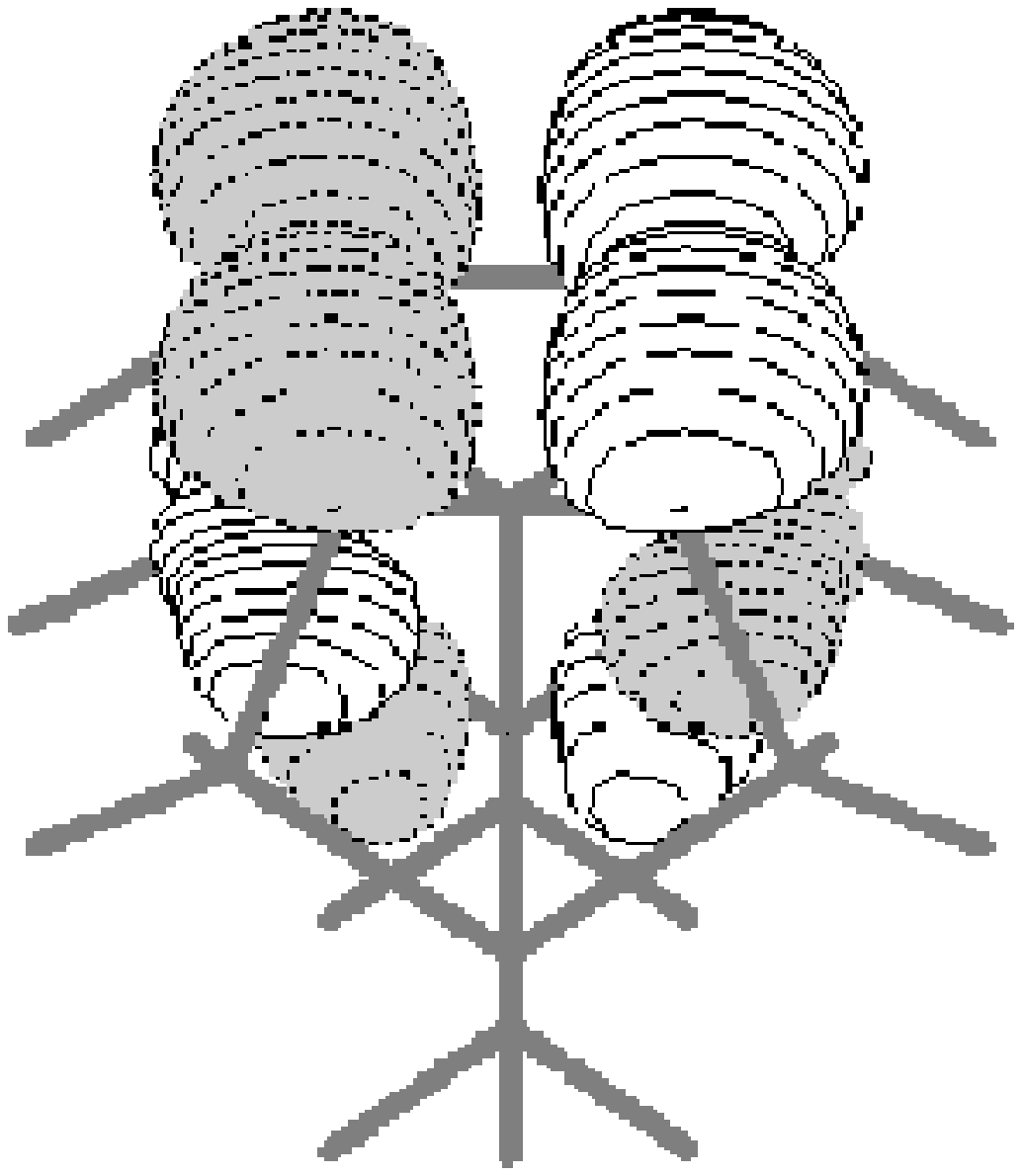,width=3.0cm}}
\vspace*{0.3cm}
\caption{Bonding (upper figure) and antibonding (lower figure) $\pi$
orbitals
for the symmetric Si$_{15}$H$_{16}$. The orbitals can be bonding (right) or
antibonding (left) between adjacent dimers.}
\label{fig3}
\end{figure}
\vspace{5pt}
\end{minipage}
Once the determinantal part of the wave function has been determined,
the parameters in the Jastrow factor are optimized within QMC using the
variance minimization method~\cite{Umrigar_vmin} and the accuracy of the
wave function is tested using variational Monte Carlo (VMC).  Reoptimizing
the determinantal part of the wave function~\cite{efp} in the presence of
the Jastrow factor did not lead to a significant improvement in the energy.
The wave function is then used in diffusion Monte Carlo (DMC),
which produces the best energy within the fixed-node approximation
(i.e. the lowest-energy state with the same nodes as the trial wave
function)~\cite{DMC}.

{\it Results and conclusions.}
Since the interplay between static and dynamical correlations is central
to this problem, we first investigate the importance of accounting for
near-degenerate molecular orbitals in the two-dimer clusters.
For the symmetric geometry, the inclusion of determinants corresponding
to the active $\pi$-orbitals (Fig.~\ref{fig3}) yields a better wave function, 
with a VMC energy which is 0.34$\pm$0.06 eV per dimer lower than the 
single-determinant energy. However, within DMC, single and multi-determinant 
wave functions yield energies which only differ by 0.04$\pm$0.02 eV per dimer.
Since the DMC energy of the symmetric reconstruction is rather
insensitive to the use of more than one determinant, it is not surprising that,
for the buckled geometry, no energy gain is obtained neither in VMC nor DMC
by using a multi-determinant wave function:
the larger HOMO-LUMO gap of the buckled cluster makes a single determinant
the best option also at the variational level.

Since a multi-determinant treatment fails to yield a significant energy
gain for the two-dimer cluster, the calculations for the Si$_{21}$H$_{20}$
cluster are performed with one determinant only.
In Table~\ref{table1}, we list the QMC results obtained for the two- and 
three-dimer clusters using a single-determinant wave function.

\noindent
\begin{minipage}{3.375in}
\begin{table}[hbt]
\caption{VMC and DMC energy differences between the symmetric ($E_{\rm sym}$) and 
buckled ($E_{\rm buck}$) reconstructions of the Si$_{15}$H$_{16}$ and 
Si$_{21}$H$_{20}$ clusters.  $\Delta E=E_{\rm sym}-E_{\rm buck}$ and
the numbers in parenthesis are the statistical errors on the last two
figures. Energies are in eV.}
\label{table1}
\begin{tabular}{lccc}
& & $\Delta E$ & $\Delta E$/dimer \\[.1ex]
\hline\\[-2ex]
Si$_{15}$H$_{16}$ & VMC    & 0.53(13) & 0.27(06)\\
                  & DMC    & 0.06(04) & 0.03(02)\\[1.5ex]
Si$_{21}$H$_{20}$ & VMC    & 0.70(14) & 0.23(05)\\
                  & DMC    & 0.34(06) & 0.11(02)\\
\end{tabular}
\end{table}
\end{minipage}

For Si$_{15}$H$_{16}$, the symmetric and buckled reconstructions are 
energetically very close as in the case of the one-dimer cluster. With a 
DMC energy difference per dimer of 0.03$\pm$0.02 eV, we cannot establish
which reconstruction is more favorable for this cluster size.
However, for Si$_{21}$H$_{20}$, the DMC energy gain per dimer in favor
of buckling increases substantially to 0.11$\pm$0.02 eV. This is in good
agreement with the trends established within DFT that a one- or a two-dimer
cluster fails to model the surface accurately~\cite{Penev}.

In Fig.~\ref{fig4}, we summarize the energy differences between 
the symmetric and the buckled reconstruction obtained within the 
local-density approximation (LDA), several generalized gradient 
approximations (GGA), B3LYP and DMC.
We also include results for Si$_9$H$_{12}$ and the slab geometry~\cite{Penev}.
DFT functionals predict that the buckled cluster is favored by 0.07-0.13 and
0.15-0.20 eV per dimer for Si$_{15}$H$_{16}$ and Si$_{21}$H$_{20}$,
respectively.  Even though the DMC energy differences are noticeably
smaller than in DFT, they clearly indicate that buckling is energetically 
more favorable.
The DMC results also highlight the importance of dimer-dimer interactions: 
Si$_{15}$H$_{16}$ behaves still very much like the single-dimer
cluster and a Si$_{21}$H$_{20}$ cluster is needed to be able to
resolve the energy gain in favor of the buckled geometry.

\noindent
\begin{minipage}{3.375in}
\begin{figure}[hbt]
\centerline{\psfig{figure=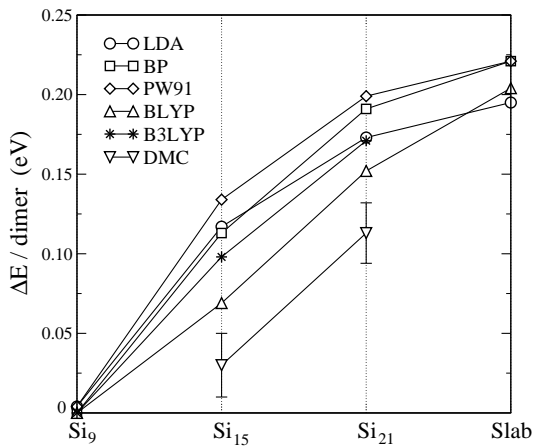,width=8.0cm,angle=-90}}
\vspace*{0.3cm}
\caption{Energy differences per dimer for the symmetric and buckled
reconstructions ($\Delta E=E_{\rm sym}-E_{\rm buck}$) obtained
with LDA, various GGA functionals, and DMC.}
\label{fig4}
\end{figure}
\vspace{1ex}
\end{minipage}
In this Letter, we presented accurate QMC calculations of the energies
of the buckled and symmetric reconstruction of the Si(100) surface. Using
large clusters, we determined that the buckled geometry is lower in energy.
Thus, recent speculations fueled by low-temperature STM experiments
that electron correlations at low temperatures would favor symmetric dimers
can be ruled out.
Furthermore, our calculations show that electronic interactions
between adjacent Si dimers in a row are important to obtain the buckled
ground state, and large clusters must be used to adequately 
model the Si(100) surface.
While DFT-LDA/GGA calculations tend to overestimate the energy gain due 
to buckling, the correct trend with cluster size is 
reproduced already at the LDA/GGA level of theory.

{\it Acknowledgements.}
We thank C. Umrigar and S. Fahy for useful discussions, and E. Shirley 
for the use of his code to generate the HF pseudopotential.
This work is supported by Enterprise Ireland Grant SC/99/242 and the Irish
Higher Education Authority.

\end{multicols}


\begin{references}
\bibitem{basis}
The Gaussian basis sets are (10$s$10$p$1$d$)/[3$s$3$p$1$d$] for silicon
and (10$s$1$p$)/[3$s$1$p$] for hydrogen, and are optimized
at the HF level for the Si$_9$H$_{12}$ cluster.
\end{references}
\end{document}